\documentclass[useAMS,usenatbib]{mn2e}

\usepackage{amsmath}

\usepackage{graphicx}

\usepackage{multirow}

\title[Non-convex model of 90 Antiope obtained with SAGE]{A new non-convex model of the binary asteroid 90 Antiope obtained with the SAGE modelling technique}
\author[P. Bartczak; T. Micha{\l}owski; T. Santana-Ros and G. Dudzi{\'n}ski]{P. Bartczak$^{1}$\thanks{E-mail:
przebar@amu.edu.pl}; T. Micha{\l}owski$^{1}$; T. Santana-Ros$^{1}$ and G. Dudzi{\'n}ski$^{1}$\\
$^{1} $Astronomical Observatory Institute, Faculty of Physics, Adam Mickiewicz University, S{\l}oneczna 36, 60-286 Pozna{\'n}, Poland}

\begin{document}

\date{Accepted 2014 June 23.  Received 2014 June 11; in original form 2013 December 18}

\pagerange{\pageref{firstpage}--\pageref{lastpage}} \pubyear{2013}

\maketitle

\label{firstpage}

\begin{abstract}

We present a new non-convex model of the 90 Antiope binary asteroid, derived with a modified version of the SAGE (Shaping Asteroids with Genetic Evolution) method using disk-integrated photometry only. A new variant of the SAGE algorithm capable of deriving models of binary systems is described. The model of 90 Antiope confirms the system's pole solution ($\lambda=199^{\circ}$, $\beta=38^{\circ}$, $\sigma=\pm5^{\circ}$) and the orbital period ($16.505046 \pm 0.000005$ h). A comparison between the stellar occultation chords obtained during the 2011 occultation and the projected shape solution has been used to scale the model. The resulting scaled model allowed us to obtain the equivalent radii ($R_{1}=40.4\pm0.9$ km and $R_{2}=40.2\pm0.9$ km) and the distance between the two system components ($176\pm4$ km), leading to a total system mass of ($9.14\pm0.62$)$\cdot10^{17}$ kg. The non-convex shape description of the components permitted a refined calculation of the components' volumes, leading to a density estimation of $1.67\pm0.23$ g cm$^{-3}$. The intermediate-scale features of the model may also offer new clues on the components' origin and evolution.

\end{abstract}


\section{Introduction}

The first asteroid satellite was discovered during the Galileo spacecraft encounter with the asteroid 243 Ida in 1993 \citep{be}. The following years have revealed discoveries of binary systems among the near-Earth objects, Main-Belt Asteroids, Mars-crossers, Trojans and Trans-Neptunian Objects. \cite{prav} and \cite{des08} introduced a simple division of multiple asteroid systems into: 1) \textit{large asteroids with small satellites}; 2) \textit{similar size and synchronous double asteroids}; 3) \textit{small asynchronous systems}; 4) \textit{contact-binary asteroids}; and 5) \textit{small wide binaries}.

The asteroid 90 Antiope was the first doubly synchronous system discovered with ground-based observations, using direct imaging obtained in August 2000 with the Keck Adaptive Optics (AO) system \citep{merline}. This disk-resolved images showed a binary system with similar-sized components of 85 km diameter, separated by 170 km. The orbital period of the components was found to be 16.5 h. The mass of these two components was also determined leading to a bulk density of $1.3$ g cm$^{-3}$.

\citet{des07} reported an extensively campaign of AO observations of Antiope carried out on 26 nights in 2003-2005 (mainly February -- March 2004) using the systems at Yepun-VLT (ESO, Chile) and Keck telescope at Mauna Kea (Hawaii, USA). With all available lightcurve data and AO observations they determined the physical and orbital model of the Antiope binary system using Roche ellipsoids as shape solutions. The lengths of semimajor axes of the components were determined to be $46.5 \times 43.5 \times 41.8$ km and $44.7 \times 41.4 \times 39.8$ km. The calculated orbital separation of these two components of $171 \pm 1$ km and the orbital period of $16.5051 \pm 0.0001$ h were obtained. The ecliptic coordinates of the pole of the system were determined as $\lambda$ = $200 \pm 2^{\circ}$ and $\beta = 38 \pm 2^{\circ}$. A total mass of ($8.28\pm0.22$)$\cdot10^{17}$ kg and a bulk density of $1.25 \pm 0.05$ g cm$^{-3}$ were determined, leading to a macro-porosity of 30 per cent.

The rotational lightcurves of this asteroid showed some asymmetries that could not be explained using ellipsoidal shape models. To account for that \citet{des09} introduced a large-scale depression located on one of the components. This giant crater of about 68 km in diameter had its center located at $145 \pm 5^{\circ}$ in longitude and $40 \pm 5^{\circ}$ in latitude on the trailing side of the component. This modification could better explain the observed light variations and lead to a slight modification of the bulk density to $1.28 \pm 0.04$ g cm$^{-3}$. The authors claimed that the crater could be a result of a collision between a 100 km sized proto-Antiope with another body beloging to the Themis family. The impactor body should have a diameter greather than 17 km and an impact velocity in the range of 1-4 km s$^{-1}$. Such event had a 50 per cent probability to have occurred over the age of the Themis family.

In February 2009 \citet{mar} obtained spectra of both components of Antiope from 1.1 to 2.4 $\mu$m. They turned out to be very similar, their slopes being in agreement with C-type asteroids. This indicated that both bodies were formed at the same time and from the same material. This confirmed that the system could be the result of the breakup of a proto-Antiope rubble-pile as suggested by \citet{des09}. (2009).

On 19 July 2011, the double asteroid 90 Antiope occulted the 6.7 mag LQ Aqr star. The path of this event was predicted to be visible mostly in the northern California. The stellar occultation was positively recorded by 43 stations \citep{colas} and a large scale topographical irregularity was reconstructed from the occultation timings on the southern component of the system. Its position was calculated at longitude 315$^{\circ}$ and latitude -40$^{\circ}$.

So far, the \textit{in situ} observations obtained with space missions (such as the 253 Mathilde and 433 Eros imaging collected with the NEAR Shoemaker spacecraft) revealed asteroids with concavities and topographical depressions which diverge from a convex shape representation. For this reason it is necessary to investigate methods capable of deriving non-convex models which shall provide a closer view on the real asteroids' shapes. We present a new version of the SAGE (Shaping Asteroids with Genetic Evolution) method for determining a non-convex shape model of 90 Antiope. The derived model is compared with the previous results presented in \citet{des09}. We have scaled the model projecting its shape to the star occultation silhouettes obtained in July 2011 \citep{colas}. The model permitted a refined calculation of the components' volumes leading to a higher density estimation.

\section{Asteroid shape}

\subsection{Defining the two bodies}

\begin{figure}

\centering

\includegraphics[width=85mm]{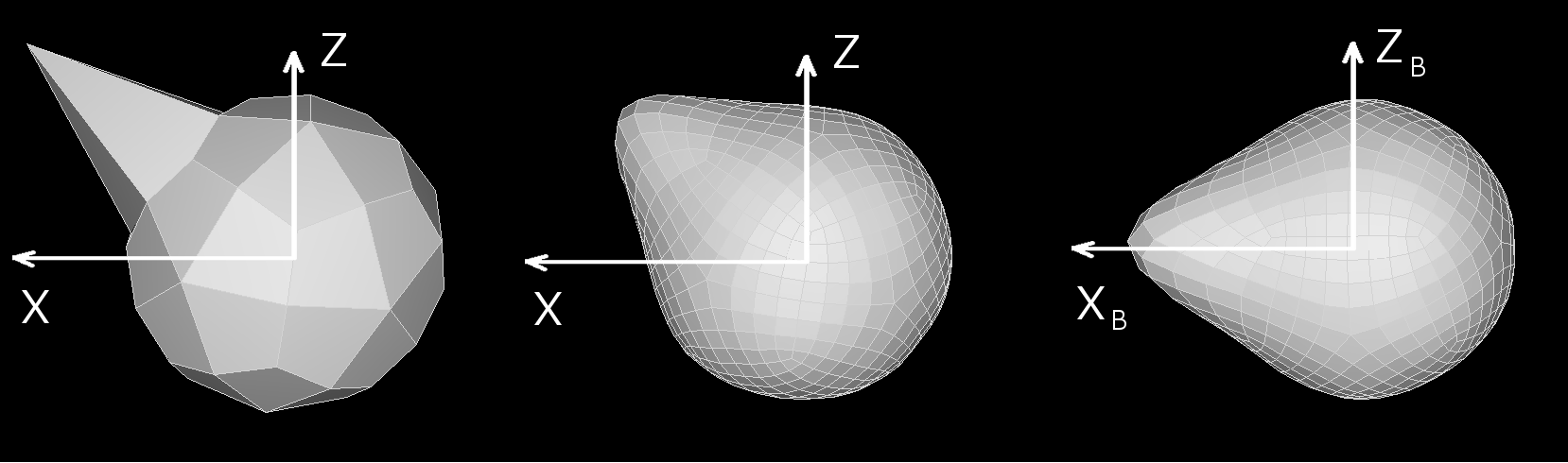}

\caption{An example of the procedure followed by the algorithm to define each component. The first shape on the left is the generator shape, i.e. a starting mesh with 62 vertices. The one in the middle is the result of applying the Catmull-Clark subdivision method to the first shape. The last step is shown on the right, where the body is oriented so the Z-axis coincides with the principal axis of rotation of the body and the origin of the body frame is translated to the position of the centre of mass.}

\label{pierwszy}

\end{figure}

\begin{figure}

\centering

\includegraphics[width=85mm]{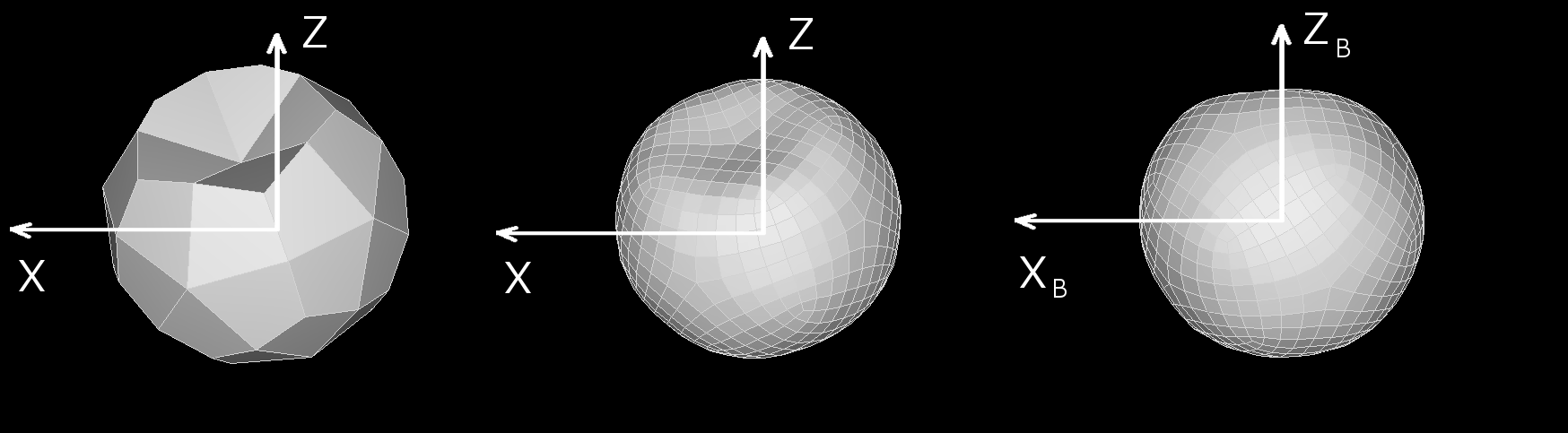}

\caption{The same procedure described in Fig.1 but for a body with a concavity.}

\label{drugie}

\end{figure}

The method starts with defining two independent components, each described by 62 vectors uniformly distributed in space and with a common origin (geometric center of the body). These vectors define a mesh of 62 vertices called a generator shape. The directions of the vectors are fixed, while their lengths evolve during the minimisation process. Once the generator shape is generated, the Catmull-Clark subdivision method is applied \citep{cat78}, resulting in a smoother shape representation of the two components, described by a higher resolution mesh. The center of mass is then calculated for each object, considering the components to be homogeneous with uniform density ${\rho}$ and total mass $M={\rho}V$. The rotational inertia of each component is characterized by its inertia tensor $I_{j,k}$, so the inertia tensor relative to the center of mass is used in order to find the bodies principal moments and axes. Finally, the main reference axes are redirected so the Z-axis coincides with the principal axis of rotation of each component. We will henceforth call this system the body frame and its axes are $X_{B}$, $Y_{B}$ and $Z_{B}$. An example of the described process can be seen for two different asteroid shapes in Fig.~\ref{pierwszy} and Fig.~\ref{drugie}.

\subsection{System construction}

\begin{figure}

\centering

\includegraphics[width=85mm]{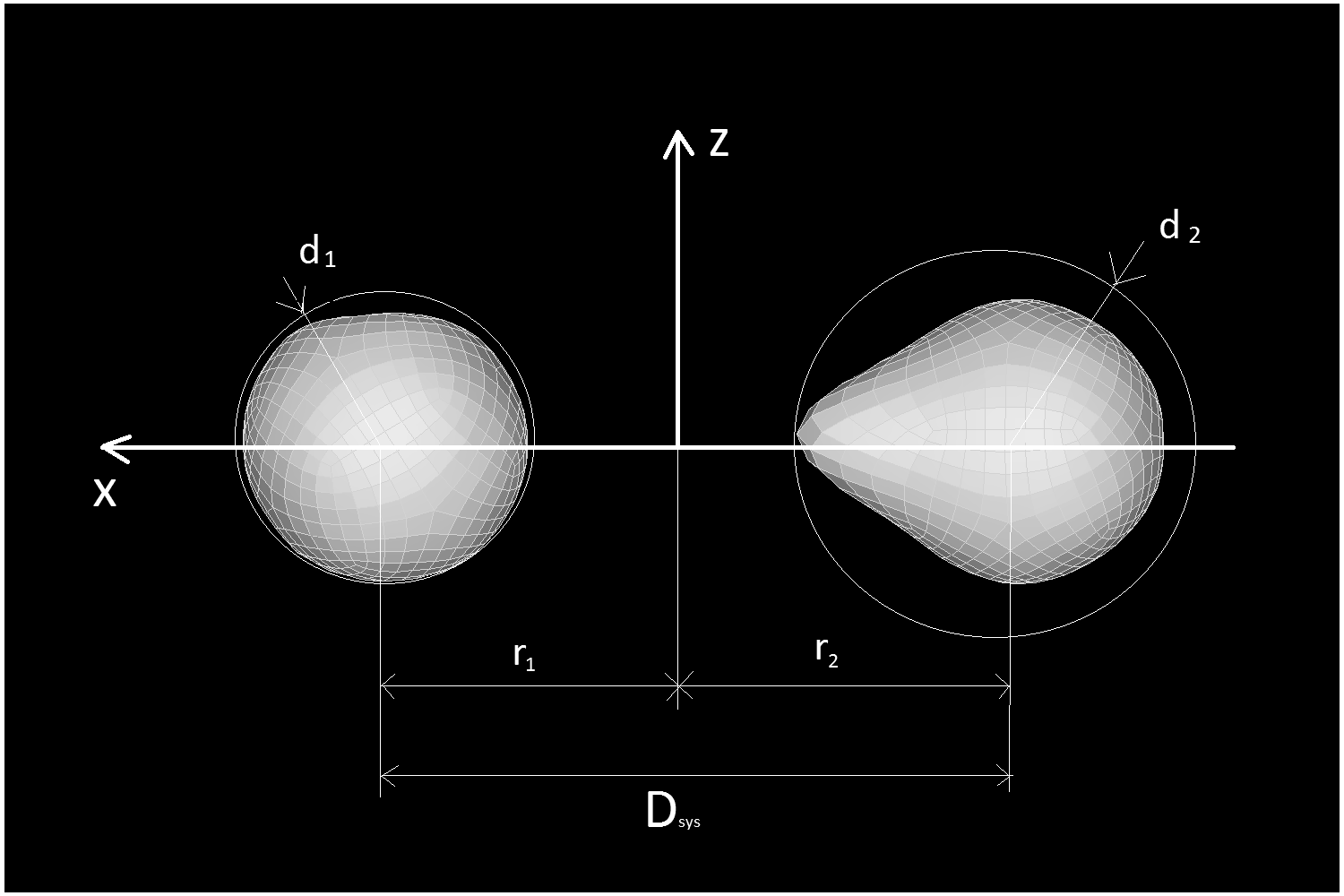}

\caption{Representation of the resulting system configuration after following the described procedures. $D_{sys}$ and $d_{1}/d_{2}$ are the parameters used to relate the two bodies during the minimization process, while $\bmath{r}_1$ and $\bmath{r}_2$ are the positions of the body centers of mass with respect to the system's center of mass.}

\label{uklad}

\end{figure}

We define a Cartesian coordinate system of axes, with the X-axis passing through the center of mass of the two objects and coincident with the body frame $X_{B}$ axis of each component. We can then place one object on the positive part of the X-axis and the other on the negative, where the origin of the system is the position of the center of mass of the whole system ($\bmath{R}_{sys}$). Moreover, we define the distance between the center of mass of each asteroid ($\bmath{r}_{1},\bmath{r}_{2}$) along the X-axis as $D_{sys}$. For each component we also define $d_{1}$ and $d_{2}$ as the norms of the vectors from the center of mass of the body to the circumscribed circle over the projected shape of the body, and $d_{1}/d_{2}$ describes the size ratio between the two components. A schema of the defined binary system can be seen in Fig.~\ref{uklad}.

These last values, $D_{sys}$ and $d_{1}/d_{2}$, are used to define the relation between the two components and will be used as parameters during the minimisation procedure, together with the 62 vertices of each body and the pole orientation of the system. An eccentrict orbit would produce a shift in the minima position between observations obtained during different apparitions. As we have not observed such phenomena, the system's orbit is considered to be circular. Therefore, the parameters to describe the system include:

\begin{itemize}

\item 62 vertex distances from the center of the first body (generator shape 1)

\item 62 vertex distances from the center of the second body (generator shape 2)

\item Distance between the center of masses ($D_{sys}$)

\item Size ratio between the two components ($d_{1}/d_{2}$) 

\item Pole solution (${\lambda},{\beta}$)

\end{itemize}
Two additional parameters (rotational period $p$ and the rotation angle of the body at zero phase ${\phi}$) are calculated by finding the best period solution scanning along a given interval.

\subsection{Center of mass, moment of inertia and rotation matrix}

In order to calculate the moments of inertia, we use the formulae described by \citet{dobro}. We assume that the components are homogeneous, with uniform density $\rho$ and total mass $M=\rho V$. Then each simplex is also homogeneous, with mass $\Delta M = \rho \Delta V$. To find the center of mass $\bmath{R}$ of each component, recall that its moment of mass $M\bmath{R}$ is just the sum of the mass moments $\Delta M$ $\Delta \bmath{R}$ of all the simplices. Therefore, the centre of mass location is given by the vector $\bmath{R}$
\begin{eqnarray}
\bmath{R} = \sum \frac{\Delta M \Delta \bmath{R}}{M} = \sum \frac{\rho \Delta V \Delta \bmath{R}}{\rho V} = 
\sum \frac{\Delta V \Delta \bmath{R}}{V},
\end{eqnarray}
and the origin of the system of axes can be translated to obtain $\bmath{R}=0$.

The rotational inertia of a rigid body may be characterized by its inertia tensor
\begin{eqnarray}
I_{j,k} =
\begin{bmatrix}
I_{x,x} & I_{x,y} & I_{x,z} \\
I_{x,y} & I_{y,y} & I_{y,z} \\  
I_{x,z} & I_{y,z} & I_{z,z} 
\end{bmatrix}.
\end{eqnarray}			
and in the given situation, $I_{j,k}$ is relative to the center of mass.

The attitude matrix is usually expressed in terms of the 3-1-3 set of the Euler angles:
rotation angle $\psi$, nutation angle $\theta$, and precession angle $\varphi$ \citep{gold}. Although the Euler angles are quite useful in describing rotation, they also posses serious drawbacks: they become undetermined for $\theta=0$ or $\theta=\pi$, and the elements of the attitude matrix depend on trigonometric functions of the angles. The latter property implies that their use in numerical integration is rather costly. In these circumstances, we prefer to use the Euler parameters \citep{gold}. Although the vector of the Euler parameters $\bmath{q} = (q_0,\,q_1,\,q_2,\,q_3)^\mathrm{T}$ consists of four elements (one more variable, compared to the Euler angles), the elements $M_{i,j}$ of the attitude matrix are easily expressible in terms of $\bmath{q}$  
\begin{eqnarray}
M_{1,1} & = & q_0^2+q_1^2-q_2^2-q_3^2, \nonumber \\
M_{1,2} & = &  2\,(q_1\,q_2+q_0\,q_3), \nonumber \\
M_{1,3} & = & -2\,(q_0\,q_2-q_1\,q_3), \nonumber \\
M_{2,1} & = & 2\,(q_1\,q_2-q_0\,q_3), \nonumber \\
M_{2,2} & = &  q_0^2-q_1^2+q_2^2-q_3^2, \\
M_{2,3}  & = &  2\,(q_2\,q_3+q_0\,q_1), \nonumber \\
M_{3,1} & = & 2\,(q_1\,q_3+q_0\,q_2), \nonumber \\
M_{3,2} & = &  2\,(q_2\,q_3-q_0\,q_1), \nonumber \\
M_{3,3} & =  & q_0^2-q_1^2-q_2^2+q_3^2, \nonumber 
\end{eqnarray}
involving only products or squares.

The relation between the Euler angles and $\bmath{q}$, often required in order to input the initial conditions, is
\begin{eqnarray}
\begin{array}{l}
q_0=\cos{\frac{\theta}{2}}\,\cos{\frac{\varphi+\psi}{2}},\\
q_1=\sin{\frac{\theta}{2}}\,\cos{\frac{\varphi-\psi}{2}},\\
q_2=\sin{\frac{\theta}{2}}\,\sin{\frac{\varphi-\psi}{2}},\\
q_3=\cos{\frac{\theta}{2}}\,\sin{\frac{\varphi+\psi}{2}}.
\end{array}
\end{eqnarray}

When the system rotates, the Euler parameters change according to the differential equations
\begin{eqnarray} \label{att}
\dot{\bmath{q}} = \frac{1}{2}
\left( \begin{array}{ccc}
-q_1 & -q_2 & -q_3 \\
q_0 & -q_3 & q_2 \\
q_3 & q_0 & -q_1 \\
-q_2 & q_1 & q_0 \\
\end{array} \right)\,\bmath{\Omega},
\end{eqnarray}
where $\bmath{\Omega}=(\Omega_1,\,\Omega_2,\,\Omega_3)^\mathrm{T}$ is the angular rate vector in the body frame. In the case of a principal axis rotation, $\Omega_1=0$, $\Omega_2=0$ and $\Omega_3=2\pi/p=const$. Except for the  components of $\bmath{q}$, subscripts $1$, $2$, $3$ refer to the axes $X_{B}$, $Y_{B}$, and $Z_{B}$ respectively.

\subsection{Generation of synthetic light curves}

\begin{figure}

\centering

\includegraphics[width=60mm]{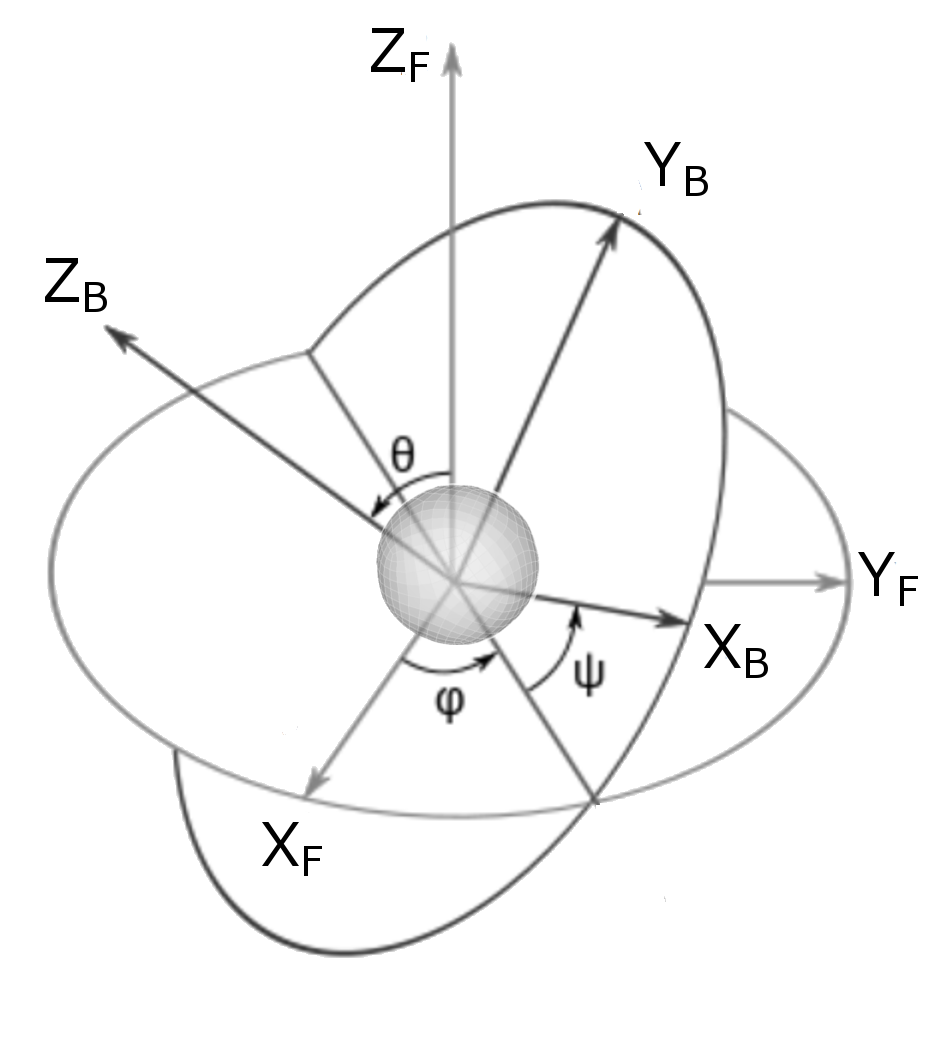}

\caption{This scheme illustrates the Euler angles ($\psi$,$\theta$,$\varphi$), relating the fixed frame (F) to the body frame (B).}

\label{fig:euler}

\end{figure}

In order to generate a synthetic picture of the components as seen on a hypothetical CCD image, we introduce a new Cartesian coordinate system of axes ($X_{F}$,$Y_{F}$,$Z_{F}$) that we call the fixed frame. This system of axes is related to the body frame using the attitude matrix (an illustration can be seen in Fig.~\ref{fig:euler}).

The fixed frame is a heliocentric system, where we can define $\bmath{S}_A$ and $\bmath{S}_E$ as the vectors in a given moment of time towards the asteroid and the Earth respectively. Those vectors can be either calculated using Keplerian orbits and the analytical expressions described in \citet{soma} or can be obtained by an ephemeris computation service such as Horizons\footnote{http://ssd.jpl.nasa.gov/horizons.cgi}.

Additionally, we describe the spin axis using the angles $\psi$ and $\theta$, while we use $\varphi$ to characterize the system rotation. We also define the vectors $\bmath{A}_E$ and $\bmath{A}_S$ as the vectors towards the Earth and the Sun from the system frame (i.e. the Earth and Sun asterocentric positions), and its transformation to the fixed frame became straighforward using the $M_{i,j}$ attitude matrix:
\begin{eqnarray}
\begin{array}{l}
\bmath{A}_E = M_{i,j} \bmath{S}_E,\\
\bmath{A}_S = M_{i,j} \bmath{S}_A.
\end{array}
\end{eqnarray}
As a mean to calculate the illumination of the components, we define a square Virtual CCD Frame. If $\bmath{rs}$ is the vector from the system's center of mass $\bmath{R}_{sys}$, to the circumscribed circle over the projected shape of the two components, then the minimum dimension of a Virtual CCD Frame which can contain the whole system is $2\,rs \times 2\,rs$ (where $rs$ is the norm of the vector $\bmath{rs}$). Each frame is divided into $N \times N$ pixels and its plane is perpendicular to the direction of $\bmath{A}_S$ and placed in an infinite distance behind the system.

Then we can make use of a Z-Buffer standard graphic method described by \citet{cat74}, projecting the components' triangular mesh onto the Virtual Frame. For each pixel in the frame we allocate a buffer with an initial value defined at an infinite distance. Before storing the number of a projected triangle of the components' mesh, the program checks the distance to the Virtual Frame. If the distance is smaller than the stored value, the new value is written. Once all the triangles are projected, we are able to determine which ones (and to what extent) are hidden by other triangles, or which are not illuminated. Finally, to calculate the brightness as seen from the observer point of view, we make the plane perpendicular to $\bmath{A}_E$ (Earth-pointing vector). The above described Z-Buffer procedure is repeated, with the brightness being now stored in the Virtual Frame. When computing the brightness, we take into account the projective shadowing and we use a linear combination of two different scattering laws, with 0.1 for the Lambert law and 0.9 for the Lommel-Seeliger law following \citet{kas}. When the procedure is done, we can compute the brightness by taking the sum of all stored values.

\section{Minimizing procedure}

The minimization procedure compares at each step the simulated brightness with the photometric observations using a $\chi^2$ test. For each single lightcurve we compute the standard deviation, which allows to obtain a normalized $\chi^2$ that takes into account the quality of the given observation. The method has one main drawback: it is high CPU demanding. For this reason it is essential to make use of a CPU cluster if we want to obtain a result in a reasonable period of time. Therefore, we used the Pozna{\'n}'s observatory cluster, that consist of 27 workstations equipped with a 6 core AMD processor (3 GHz).

\begin{figure}

\centering

\includegraphics[width=85mm]{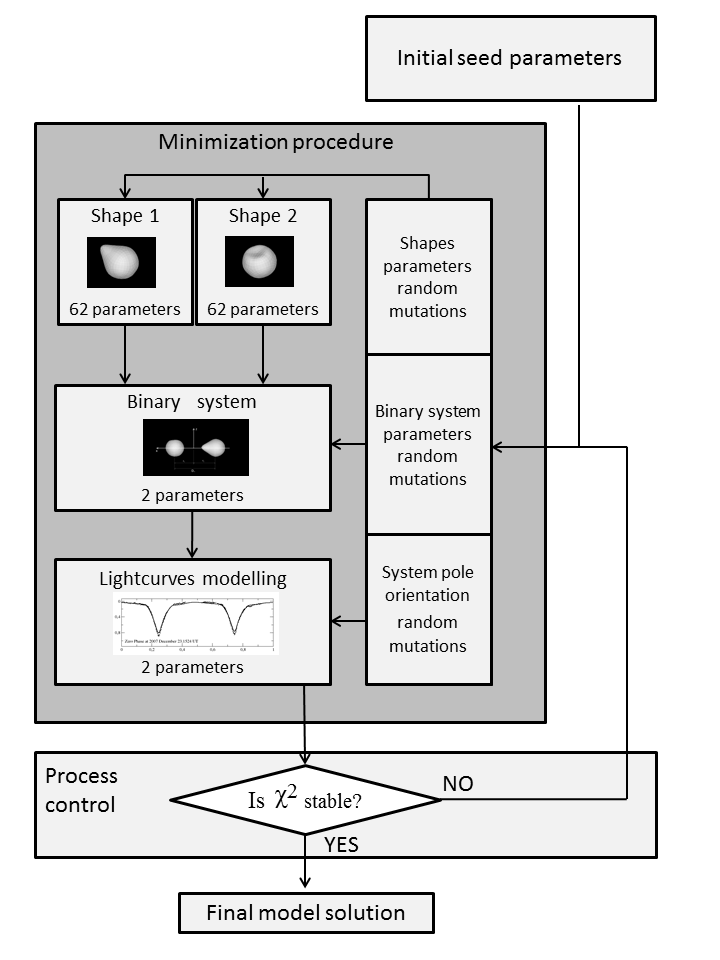}

\caption{Process flow diagram which depicts the loop sequence for the minimization algorithm. The loop ends when the minimization residual is not significantly changing during few modelling attemps. The full process have to be then restarted, feeding the system with new seed parameters. The routine is repeated and a family of similar models are obtained. The best absolute fit to lightcurves become our formal solution.}

\label{diagram}

\end{figure}

The minimising procedure consists of two modules summarized in Fig.~\ref{diagram}. The initial seed of the loop will be the sphere-like shapes of the two asteroids and a random orientation of the system. The first module randomly mutates the model parameters (the mesh of vertices, the spin parameters and the system parameters). The best trial solution found becomes the seed for the consecutive random evolution and the process is cyclically repeated. The second module (active balance) allocates different weights to the observations, calculating $\chi^2$ for each lightcurve, and giving greater weight to the observation with the worst fit. This procedure allows the method to bypass local minima. Obviously any erroneous observations (bad reduction, wrong time recorded, etc ...) must be removed previously in order not to interfere the process.

Therefore, the method tries to find the solution that best fits all the observations, and the results of the modeling are the parameters of the smallest $\chi^2$ found. The evolution of the model parameters is controlled by the genetic algorithm routines. In order to be sure that the evolution is not trapped in a local minimum, we repeat the fitting process several times, starting from the same initial spherical shapes. This generates a family of solutions, similar in global terms. The formal solution will be the family member that stabilizes in a lowest $\chi^2$. If the family members are not converging to a similar solution, it would be a clear signal that there are not enough constraints to derive the model e.g. the given asteroid should be observed in the future apparitions with different observational geometries.

\section{Lightcurve data}

\begin{table*}

\begin{minipage}{170mm}

\caption{Details of the lightcurves used for our modelling of 90 Antiope. The table columns describe the observations time range, the number of observing nights, the phase angle range, the ecliptic longitudes and latitudes of the asteroid around the opposition dates, the observed eclipsing amplitudes and the references. The value of "0" for the eclipsing amplitudes means that no eclipsing events were observed for these apparitions (i.e. 2000 and 2006).}

  \label{observations}

\begin{tabular}{cccccccc}

\hline

\hline

Apparition & Time range & N$_{lc}$ & $\alpha$ $[^{\circ}]$ & $\lambda$ $[^{\circ}]$ & $\beta$ $[^{\circ}]$ & Eclipsing  & Reference\\

 &  &  &  &  &  & amplitude [mag] &\\

\hline

1 & Dec 1996 & 4 & 7.7 - 9.8 & 120 & 1.9 & 0.56 & \citet{han}\\

2 & Sep-Nov 2000   & 14  & 3.0 - 17.7 & 355 & -2.7 & 0 & \citet{mich01}\\

3 & Oct 2001 - Feb 2002 & 26 & 0.6 - 15.0 & 80 & 0.5 & 0.05 - 0.12 & \citet{mich02}\\

4 & Dec 2002 - Apr 2003 & 31 & 0.9 - 14.7 & 138 & 2.8 & 0.00 - 0.05 & \citet{mich04}\\

5 & May - Nov 2005 & 38 & 1.3 - 21.2 & 285 & -2.1 & 0.45 - 0.75 & \citet{des07}\\

6 & Aug - Sep 2006 & 6 & 14.1 - 18.7 & 42 & -2.0 & 0 & \citet{veli}\\

7 & Nov 2007 - Mar 2008 & 38 & 0.5 - 15.4 & 104 & 1.7 & 0.70 - 0.75 & \citet{des09}\\

\hline

\end{tabular}

\end{minipage}

\end{table*}

There are lightcurves of 90 Antiope obtained during 7 apparitions (see Table \ref{observations}). During some of them (1996, 2001-02, 2002-03, 2005, 2007-08) the asteroid displayed a two-component lightcurve with each of them showing the same period of 16.505 hrs, consistent with the value presented above. The first asymmetrical component was associated with the rotation of two non-spherical bodies giving a so-called rotational lightcurve, while the second one, consisting of two sharp minima, was due to mutual events in the binary system (eclipsing lightcurve). The observations showed that both period components were equal, meaning that the Antiope binary system is synchronous. The eclipsing minima were always a half period apart indicating circular orbits of both bodies in the binary system. The values of the eclipsing amplitude were in the range of 0.00-0.75 mag during different apparitions, so data have been collected in different system geometries, i.e. without eclipses, with partial and full events. Details of the lightcurve data used for our modelling of 90 Antiope are shown in Table~\ref{observations}.

\section{Model of 90 Antiope}

\begin{figure*}

\centering

\fbox{\includegraphics[width=170mm]{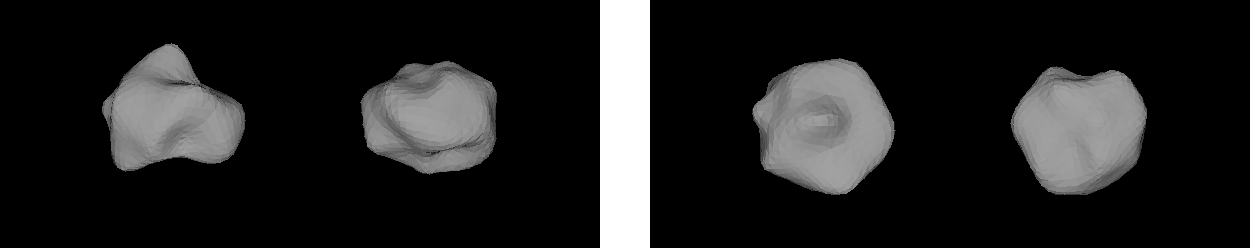}}

\caption{Two different spatial views of the non-convex model for 90 Antiope shown at equatorial viewing (on the \textit{left}), and the pole-on view on the \textit{right}.}

\label{antiope}

\end{figure*}

\begin{figure*}

\centering

\fbox{\includegraphics[width=150mm]{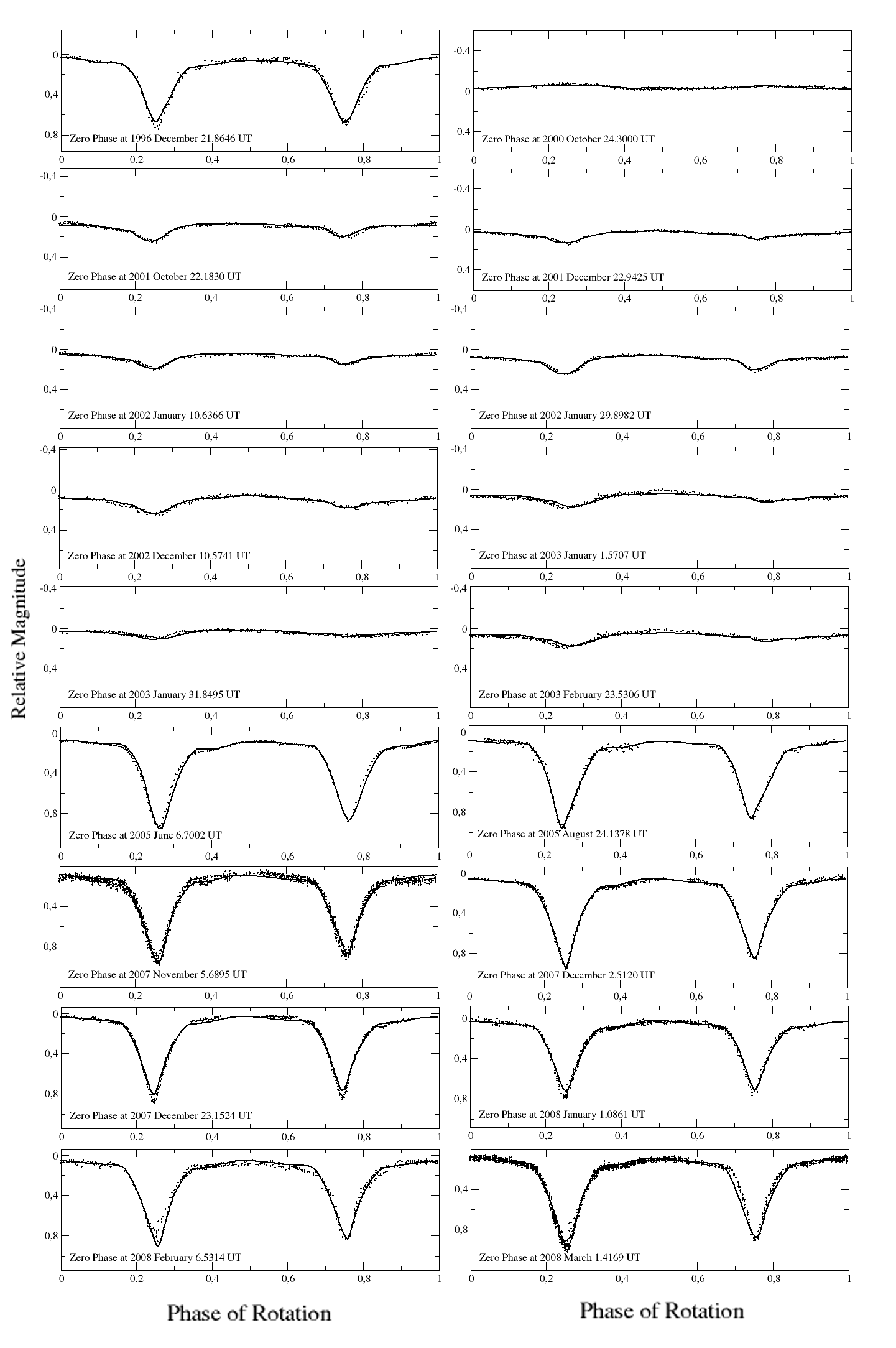}}

\caption{Antiope model fit to lightcurves. The solid line is the synthetic brightness associated with the model solution, while the dots correspond to the photometric observations (see Table~\ref{observations} for details).}

\label{lightcurves}

\end{figure*}

Using the photometric data summarized in Table 1 and the algorithm described above we have obtained a non-convex model of 90 Antiope. The presented model of this asteroid, now publicly available on the ISAM webpage\footnote{http://isam.astro.amu.edu.pl/}, is successfully reproducing the observed lightcurves, i.e. the part associated with the rotation of the two components (including the intermediate-scale topological features) and the sharp minima due to the mutual events (see Fig.~\ref{lightcurves} and the Appendix for some examples). The model is a first-order model, including the main physical photometric effects, such as the limb-darkening, the mutual shadowing controlled by the solar phase angle and the large-scale shape effects. The system's physical parameters are summarized in Table 2.

\begin{figure*}

\centering

\setlength\fboxsep{0pt}

\setlength\fboxrule{0pt}

\fbox{\includegraphics[width=170mm]{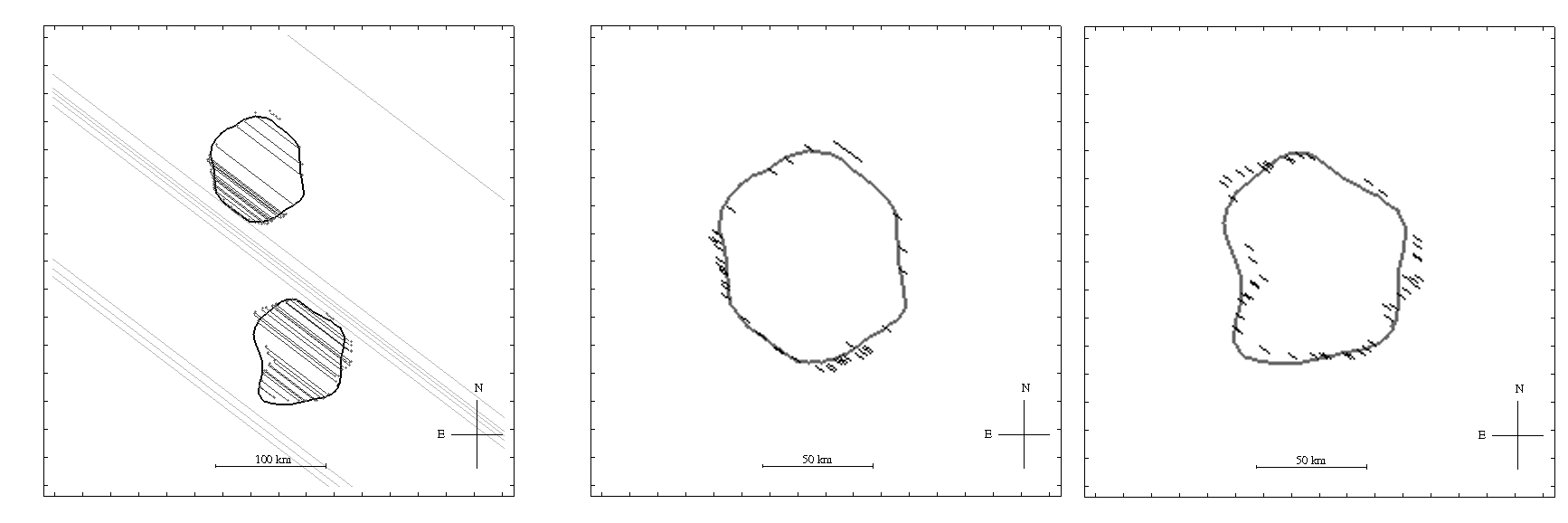}}

\caption{Left: Best fit of the stellar occultation chords obtained during the 2011 occultation \citep{colas} to the solution found for the non-convex shape model of 90 Antiope. Right: Timing uncertanties due to the unknown diameter of the occulted star (LQ Aqr).}

\label{zakrycie}

\end{figure*}

Two spatial views of the resulting shape solution are shown in Fig.~\ref{antiope}. Both components have similar dimensions but one of them (component on the \textit{left} in Fig.~\ref{antiope}) presents large-scale depressions on the trailing side, which are in accordance with the giant crater hypothesis suggested by \citet{des09}. Similar topological features have been imaged in some asteroids during spacecraft encounters. For instance, the Near Earth Asteroid Rendezvous (NEAR) spacecraft revealed five giant craters on the C-type Asteroid 253 Mathilde, four with diameters larger than the radius of Mathilde itself \citep{vever}. The composition and internal structure of Mathilde may contribute to the retention of giant craters. The bulk density of Mathilde is low (about $1.3\pm0.2$ g cm$^{-3}$), which means that the asteroid should be a rubble pile or it is composed of an extremely porous material \citep{chapman}. This porosity may dampen the propagation of shock waves through the interior of the body and may also dramatically reduce ejecta velocities. Therefore, as Antiope is also a C-type asteroid with low bulk density, this theory might also explain the formation of the large-scale depressions observed in our model.

As the presented method is based only on relative photometry, it does not allow to derive the absolute dimensions of the components and their separation. If we denote the distance between the centers of masses of these two bodies as $D_{sys}$, their relative volumes can be determinated as $V_{1}=0.050484{D_{sys}}^{3}$ and $V_{2}=0.049808{D_{sys}}^{3}$.

However, we could use the excellent results obtained during the 2011 Antiope's stellar occultation to project the shape model to the asteroid's silhouette derived from the occultation timings. As discussed by Herald (2012), the star occulted by Antiope (LQ Aquarii) is a "slow red" star (LB) without direct measurements of its diameter. The main issue during the observation reduction is thus to take into account the non negligible time needed to occult the star. \citet{colas} gave an estimation of the star's diameter of $1.7\pm0.7$ $mas$, which is equivalent to $2.2\pm0.9$ km at the asteroid level. Having this in mind, we have developed an optimization algorithm which looks for the best fit of the projected shape of the non-convex model to the occultation timings. The best fit is shown in Fig.~\ref{zakrycie}, and can be qualified as good as the average difference between the model and the occultation timings is $4$ km. As a result of this fit, we were able to scale the model, deriving the absolute physical properties of the system. The orbital separation of the two bodies was calculated to be $D_{sys}=176\pm4$ km, slightly larger than the $D_{sys}=170\pm1$ km calculated by \citet{des07} using direct imaging observations, but smaller compared to the $D_{sys}=180$ km obtained by \citet{colas} derived from the 2011 stellar occultation.

The obtained separation $D_{sys}$ allowed us to calculate the volumes of the components and their equivalent radii, i.e. the radii of spheres that have the same volumes as the non-convex models. The equivalent radii of both bodies are formally equal. Similar volumes (and equivalent radii) of such irregular bodies mean that cross sections might vary with orbital phase of the components.

\begin{table}

\caption{Orbital and physical model of 90 Antiope.}

  \label{modelresults}

\begin{tabular}{c|c}

\hline

\hline

\textbf{Parameter} & \textbf{Model result}    \\

\hline

Orbital period & $16.505046 \pm 0.000005$ h\\

Pole of the orbit: &   \\

 $\lambda$          & $199^{\circ}$ \\

 $\beta$            & $38^{\circ}$ \\

 pole position error & $\pm5^{\circ}$ \\

Separation    & $176\pm4$ km\\

Total mass    & $(9.14\pm0.62)\cdot10^{17}$ kg\\

Equivalent radii: &   \\

   $R_{1}$      & $40.4\pm0.9$ km \\

   $R_{2}$      & $40.2\pm0.9$ km\\

Bulk density  & $1.67\pm0.23$ g cm$^{-3}$\\

\hline

\end{tabular}

\end{table}

Knowing the distance between the two centers of mass and the orbital period, we calculated the total mass of the system, resulting in $(9.14\pm0.62)\cdot10^{17}$ kg. This value is slightly higher than the one reported by Descamps et al. (2007, 2009). However, the equivalent radii (and volumes) are smaller than the ones obtained in previous studies, thus the inferred bulk density of 90 Antiope resulted to be 30 per cent larger (see Table~\ref{modelresults}).

\section{Conclusions}

We derived a non-convex model of 90 Antiope which reproduces its photometric observations and fits the chords obtained during the 2011 stellar occultation \citep{colas}. The latter observations enabled us to scale the model and derive its fundamental physical properties. We have found a density 30 per cent higher than the one calculated in previous studies (\citealt{merline}, \citealt{des09}). On the other hand, the components' separation resulting from our calculations ($176\pm4$ km) is larger than the values presented in studies based on disk resolved imaging (\citealt{merline}, \citealt{des07}), but smaller compared to the value obtained with a Roche ellipsoid model \citep{colas}. The intermidate-scale features revealed by the shape solution supports the giant crater hypothesis presented by \citet{des09} and confirms that large-scale depressions are present on the trailing side of the southern component.

\section*{Acknowledgments}

This work was partially supported by grant N N203 404139 from the Polish Ministry of Science and Higher Education.

The work of TSR was carried out through the Gaia Research for European  Astronomy  Training  (GREAT-ITN)  network.  He  received  funding  from  the  European  Union  Seventh Framework Programme (FP7/2007-2013) under grant agreement No. 264895.

\onecolumn

\newpage

\appendix

\section{Model fits of 90 Antiope to observations}

\begin{figure}

\centering

\fbox{\includegraphics[width=140mm]{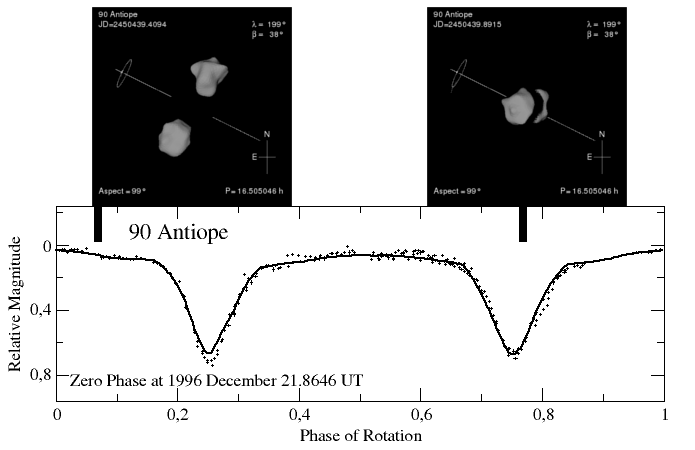}}

\caption{Antiope model fit to the observations obtained in December 1996. The solid line is the synthetic brightness associated with the model solution, while the dots correspond to the photometric observations \citep{han}. The vertical bars indicate the phase moments selected for the two snapshots of the model. The snapshots have been obtained using the ISAM, where the presented model is now publicly available.}

\label{LC0}

\end{figure}

\begin{figure}

\centering

\fbox{\includegraphics[width=140mm]{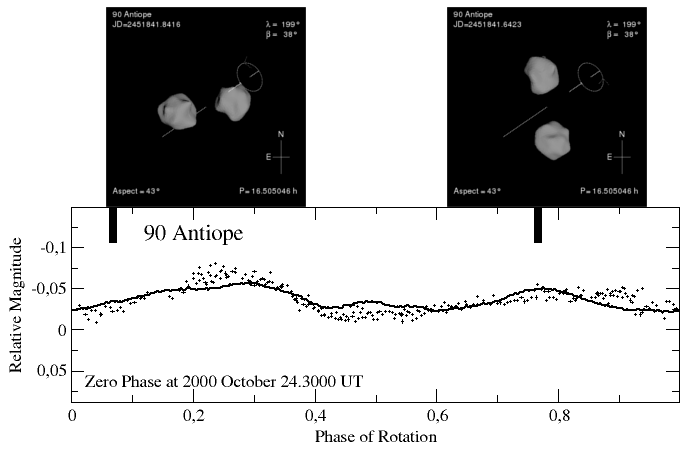}}

\caption{The same as in Fig.~\ref{LC0} but for the observations obtained in October 2000 \citep{mich01}.}

\label{LC1}

\end{figure} 

\begin{figure}

\centering

\fbox{\includegraphics[width=140mm]{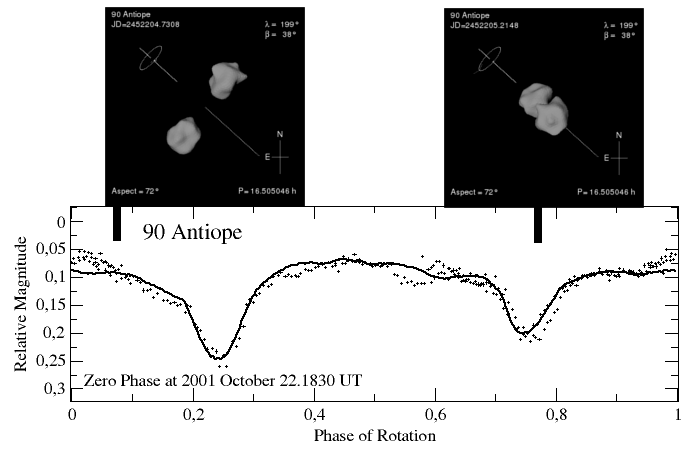}}

\caption{The same as in Fig.~\ref{LC0} but for the observations obtained in October 2001 \citep{mich02}.}

\label{LC2}

\end{figure} 

\begin{figure}

\centering

\fbox{\includegraphics[width=140mm]{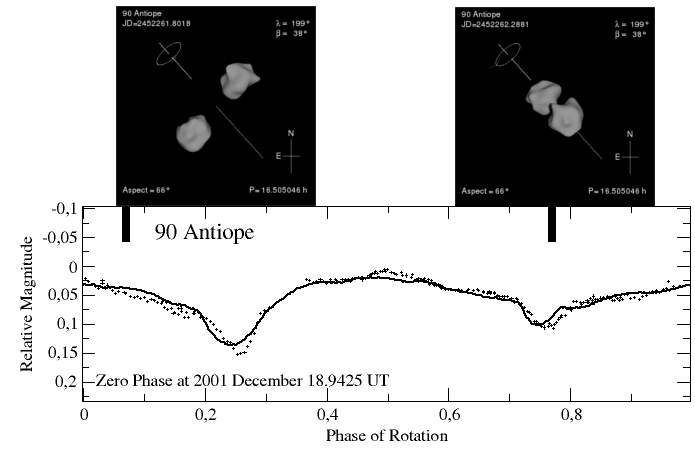}}

\caption{The same as in Fig.~\ref{LC0} but for the observations obtained in December 2001 \citep{mich02}.}

\label{LC3}

\end{figure}

\begin{figure}

\centering

\fbox{\includegraphics[width=150mm]{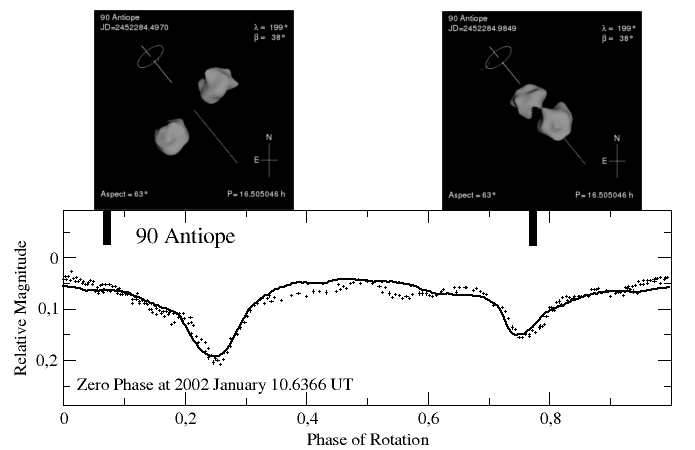}}

\caption{The same as in Fig.~\ref{LC0} but for the observations obtained in January 2002 \citep{mich02}.}

\label{LC4}

\end{figure}

\begin{figure}

\centering

\fbox{\includegraphics[width=140mm]{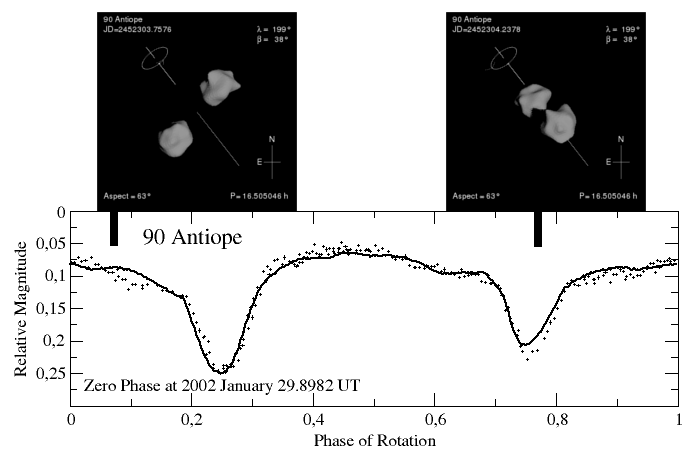}}

\caption{The same as in Fig.~\ref{LC0} but for the observations obtained in January 2002 \citep{mich02}.}

\label{LC5}

\end{figure}

\begin{figure}

\centering

\fbox{\includegraphics[width=140mm]{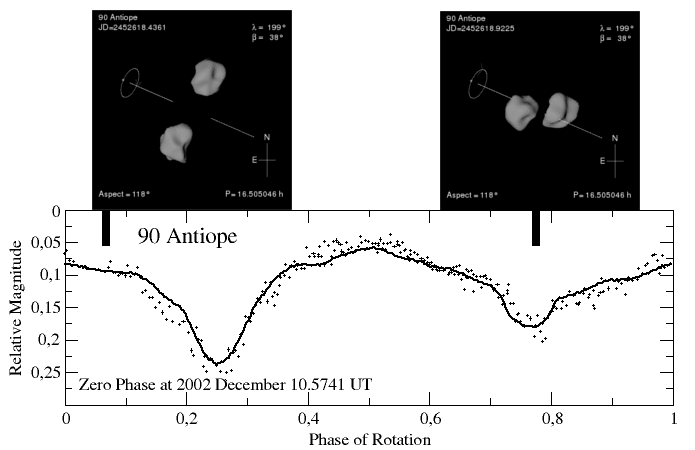}}

\caption{The same as in Fig.~\ref{LC0} but for the observations obtained in December 2002 \citep{mich04}.}

\label{LC6}

\end{figure}

\begin{figure}

\centering

\fbox{\includegraphics[width=140mm]{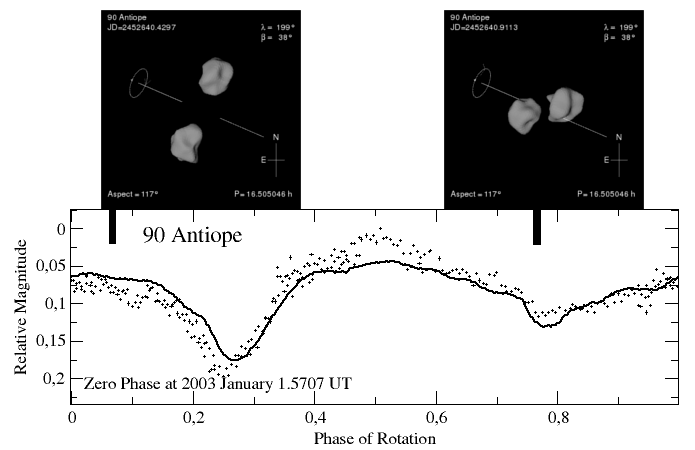}}

\caption{The same as in Fig.~\ref{LC0} but for the observations obtained in January 2003 \citep{mich04}.}

\label{LC7}

\end{figure}

\begin{figure}

\centering

\fbox{\includegraphics[width=140mm]{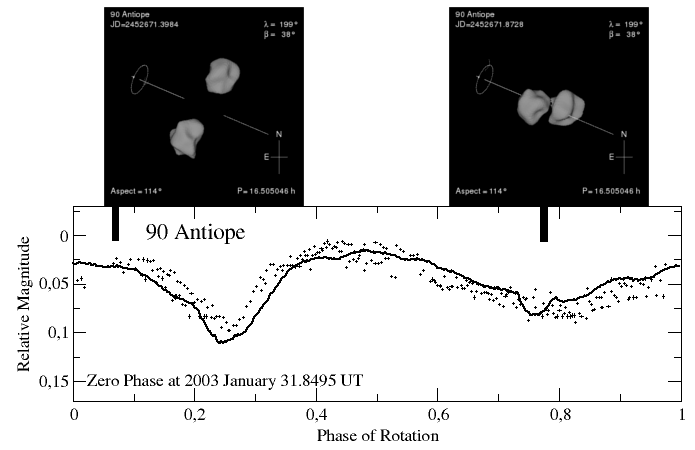}}

\caption{The same as in Fig.~\ref{LC0} but for the observations obtained in January 2003 \citep{mich04}.}

\label{LC8}

\end{figure}

\begin{figure}

\centering

\fbox{\includegraphics[width=140mm]{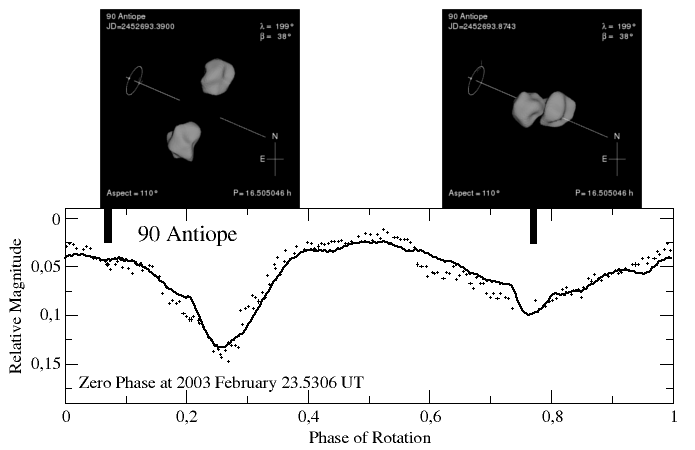}}

\caption{The same as in Fig.~\ref{LC0} but for the observations obtained in February 2003 \citep{mich04}.}

\label{LC9}

\end{figure}

\begin{figure}

\centering

\fbox{\includegraphics[width=140mm]{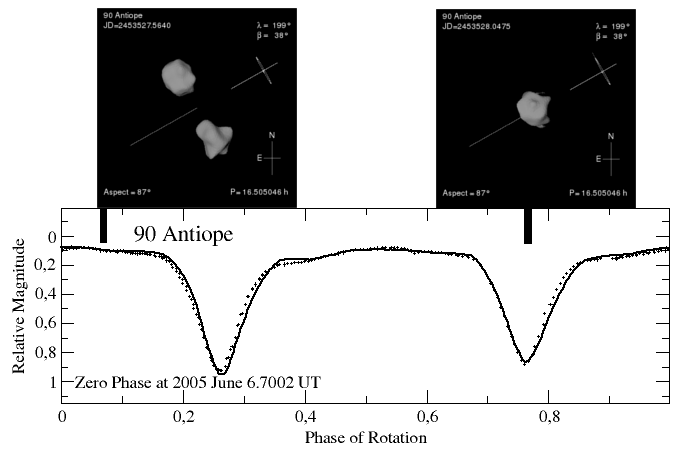}}

\caption{The same as in Fig.~\ref{LC0} but for the observations obtained in June 2005 \citep{des07}.}

\label{LC10}

\end{figure}

\begin{figure}

\centering

\fbox{\includegraphics[width=140mm]{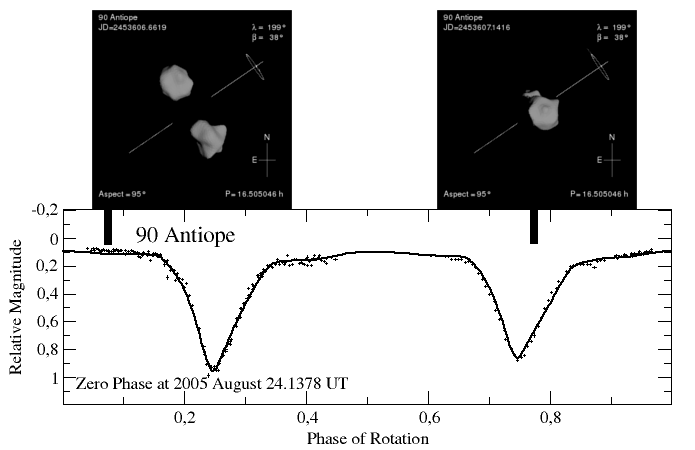}}

\caption{The same as in Fig.~\ref{LC0} but for the observations obtained in August 2005 \citep{des07}.}

\label{LC11}

\end{figure}

\begin{figure}

\centering

\fbox{\includegraphics[width=140mm]{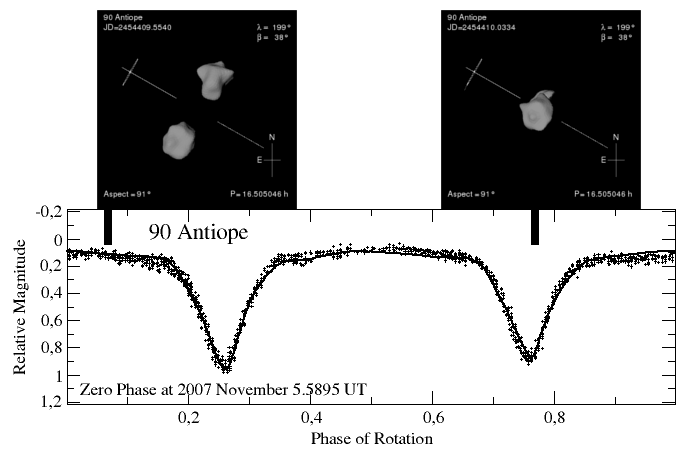}}

\caption{The same as in Fig.~\ref{LC0} but for the observations obtained in November 2007 \citep{des09}.}

\label{LC12}

\end{figure}

\begin{figure}

\centering

\fbox{\includegraphics[width=140mm]{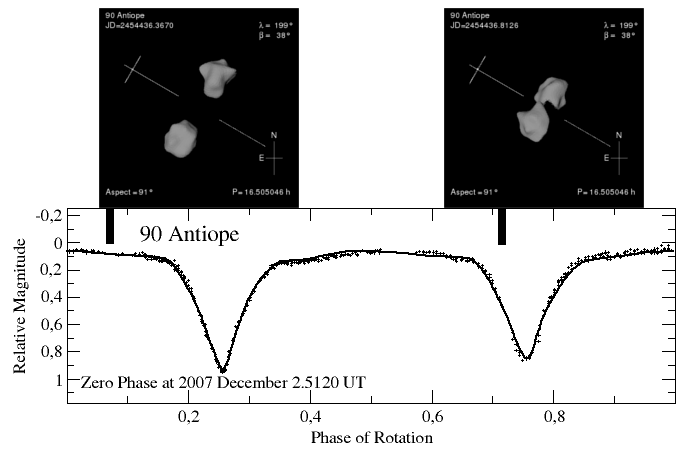}}

\caption{The same as in Fig.~\ref{LC0} but for the observations obtained in December 2007 \citep{des09}.}

\label{LC13}

\end{figure}

\begin{figure}

\centering

\fbox{\includegraphics[width=140mm]{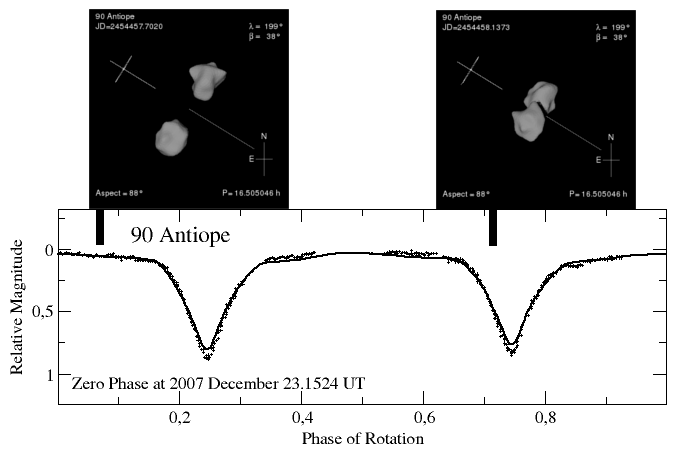}}

\caption{The same as in Fig.~\ref{LC0} but for the observations obtained in December 2007 \citep{des09}.}

\label{LC14}

\end{figure}

\begin{figure}

\centering

\fbox{\includegraphics[width=140mm]{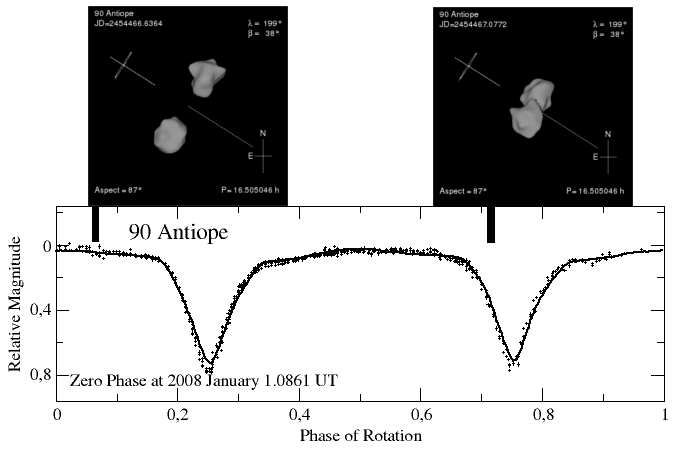}}

\caption{The same as in Fig.~\ref{LC0} but for the observations obtained in January 2008 \citep{des09}.}

\label{LC15}

\end{figure}

\begin{figure}

\centering

\fbox{\includegraphics[width=140mm]{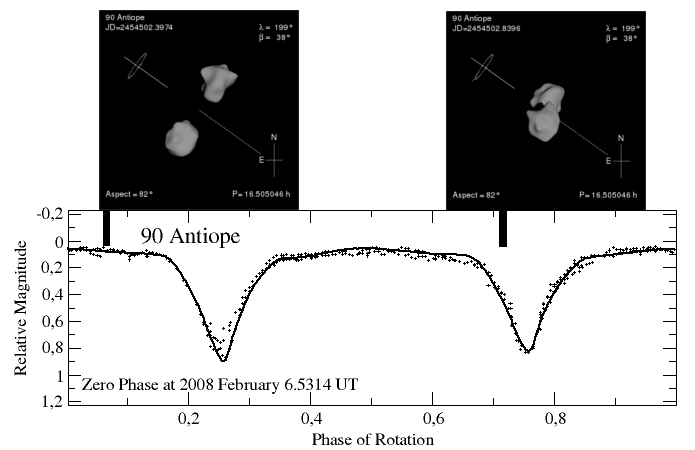}}

\caption{The same as in Fig.~\ref{LC0} but for the observations obtained in February 2008 \citep{des09}.}

\label{LC16}

\end{figure}

\begin{figure}

\centering

\fbox{\includegraphics[width=140mm]{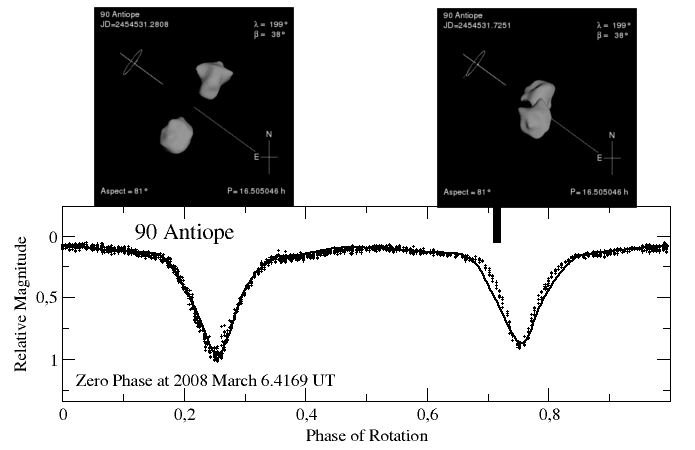}}

\caption{The same as in Fig.~\ref{LC0} but for the observations obtained in March 2008 \citep{des09}.}

\label{LC17}

\end{figure}

\label{lastpage}

\end{document}